\documentclass[11pt]{article}
\usepackage{amsfonts}
\usepackage{epsfig, graphicx}
\usepackage{amssymb}
\usepackage{mathtools}
\usepackage{bbm}
\mathtoolsset{showonlyrefs}
\usepackage{amstext}
\usepackage{amsmath,enumerate}
\usepackage{theorem}
\usepackage{algorithm}
\usepackage{algpseudocode}
\usepackage[left=1in,right=1in,top=1in,bottom=1in,nohead]{geometry}
\usepackage[active]{srcltx}
\allowdisplaybreaks[2]

\def\cA{{\cal A}}
\def\cM{{\cal M}}

\def\a{\alpha}

\def\d{\delta}
\def\D{\Delta}
\def\e{\varepsilon}
\def\f{\phi}
\def\g{\gamma}
\def\G{\Gamma}

\def\th{\theta}

\def\l{\lambda}

\def\n{\nu}

\def\r{\rho}

\def\s{\sigma}

\newcommand{\set}[1]{\left\{#1\right\}}

\newcommand{\proofstart}{{\bf Proof\hspace{2em}}}
\newcommand{\proofend}{\hspace*{\fill}\mbox{$\Box$}}

\def\E{{\sf E}}
\def\Pr{{\sf P}}

\def\cB{{\cal B}}

\newcommand{\ignore}[1]{}

\newtheorem{theorem}{Theorem}

\newtheorem{lemma}[theorem]{Lemma}

\newtheorem{remark}[theorem]{Remark}
\newtheorem{claim}[theorem]{Claim}

\newcommand{\brac}[1]{\left(#1\right)}
\newcommand{\bfrac}[2]{\brac{\frac{#1}{#2}}}

\newcommand{\beq}[2]{\begin{equation}\label{#1}{#2}\eeq}
\newcommand{\eeq}{\end{equation}}
\newcommand{\blem}[1]{\begin{lemma}\label{#1}}
\newcommand{\elem}{\end{lemma}}
\newcommand{\bth}[1]{\begin{theorem}\label{#1}}
\newcommand{\enth}{\end{theorem}}
\newcommand{\brem}[1]{\begin{remark}\label{#1}}
\newcommand{\erem}{\end{remark}}

\usepackage{color}

\author{Alan Frieze\thanks{Research supported in part by NSF Grants DMS1362785, CCF1522984 and Grant 333329 from the Simons Foundation} and Tony Johansson\thanks{Research supported in part by NSF Grant DMS1362785}\\Department of Mathematical Sciences\\Carnegie Mellon University\\Pittsburgh PA15213\\U.S.A.\\
alan@random.math.cmu.edu\\
tjohanss@andrew.cmu.edu}
\title{On the insertion time of random walk cuckoo hashing}

\date{}

\begin{document}
\maketitle
\begin{abstract}
Cuckoo Hashing is a hashing scheme invented by Pagh and Rodler \cite{cuckoo}. It uses $d\geq 2$ distinct hash functions to insert items into the hash table. It has been an open question for some time as to the expected time for Random Walk Insertion to add items. We show that if the number of hash functions $d=O(1)$ is sufficiently large, then the expected insertion time is $O(1)$ per item.
\end{abstract}
\section{Introduction}
Our motivation for this paper comes from Cuckoo Hashing (Pagh and Rodler \cite{cuckoo}). Briefly each one of $n$ items $x\in L$ has $d$ possible locations $h_1(x),h_2(x),\ldots,h_d(x)\in R$, where $d$ is typically a small constant and the $h_i$ are hash functions, typically assumed to behave as independent fully random hash functions.  (See \cite{MVad} for some justification of this assumption.) 

We assume each location can hold only one item.  Items are inserted consecutively and when an item $x$ is inserted into the table, it can be placed immediately if one of its $d$ locations is currently empty.  If not, one of the items in its $d$ locations must be displaced and moved to another of its $d$ choices to make room for $x$.  This item in turn may need to displace another item out of one of its $d$ locations.  Inserting an item may require a sequence of moves, each maintaining the invariant that each item remains in one of its $d$ potential locations, until no further evictions are needed.

We now give the formal description of the mathematical model that we use. We are given two disjoint sets $L=\set{v_1,v_2,\ldots,v_n},R=\set{w_1,w_2,\ldots,w_m}$. Each $v\in L$ independently chooses a set $N(v)$ of $d\geq 2$ uniformly random neighbors in $R$. We assume for simplicity that this selection is done with replacement. This provides us with the bipartite cuckoo graph $\G$. Cuckoo Hashing can be thought of as a simple algorithm for finding a matching $M$ of $L$ into $R$ in $\G$. In the context of hashing, if $\set{x,y}$ is an edge of $M$ then $y\in R$ is a hash value of $x\in L$.

Cuckoo Hashing constructs $M$ by defining a sequence of matchings $M_1,M_2,\ldots,M_n$, where $M_k$ is a matching of $L_k=\set{v_1,v_2,\ldots,v_k}$ into $R$. We let $\G_k$ denote the subgraph of $\G$ induced by $L_k\cup R$. We let $R_k$ denote the vertices of $R$ that are covered by $M_k$ and define the function $\f_k:L_k\to R_k$ by asserting that $M_k=\set{\set{v,\f_k(v)}:v\in L_k}$. We obtain $M_{k}$ from $M_{k-1}$ by finding an augmenting path $P_k$ in $\G_k$ from $v_{k}$ to a vertex in $\bar R_{k-1}=R\setminus R_{k-1}$. 

This augmenting path $P_k$ is obtained by a random walk. To begin we obtain $M_1$ by letting $\f_1(v_1)$ be a uniformly random member of $N(v_1)$, the neighbors of $v_1$. Having defined $M_k$ we proceed as follows: Steps 1 -- 4 constitute round $k$.

{\bf Algorithm} {\sc insert:}
\begin{enumerate}[{\bf Step 1}]
\item $x\gets v_{k}$; $M\gets M_{k-1}$;
\item {\bf If} $S_k(x)=N(x)\cap \bar R_{k-1}\neq\emptyset$ {\bf then} choose $y$ uniformly at random from $S_k(x)$ and let $M_{k}=M\cup\set{\set{x,y}}$, {\bf else}
\item Choose $y$ uniformly at random from $N(x)$;
\item $M\gets M\cup\set{\set{x,y}}\setminus\set{y,\f_{k-1}^{-1}(y)}$; $x\gets \f_{k-1}^{-1}(y)$; {\bf goto} Step 2.
\end{enumerate}
This algorithm was first discussed in the conference version of \cite{FPSS}. Our interest here is in the expected time for {\sc insert} to complete a round. Our results depend on $d$ being large. In  this case we will improve on the bounds on insertion time given in Frieze, Melsted and Mitzenmacher \cite{FMM}, Fountoulakis, Panagiotou and Steger \cite{FPS}, Fotakis, Pagh, Sanders and Spirakis \cite{FPSS}. The paper \cite{FPSS} studied the efficiency of insertion via Breadth First Search and also carried out some experiments with the random walk approach. The papers \cite{FMM} and \cite{FPS} considered insertion by random walk and proved that the expected time to complete a round can be bounded by $\log^{2+o_d(1)}n$, where $o_d(1)$ tends to zero as $d\to\infty$. The paper \cite{FPS} improved on the  space requirements in \cite{FMM}. They showed that given $\e$, their bounds hold for any $d$ large enough to give the existence of a matching w.h.p. Mitzenmacher \cite{M} gives a survey on Cuckoo Hashing and Problem 1 of the survey asks for the expected insertion time. 

Frieze and Melsted \cite{FM}, Fountoulakis and Panagiotou \cite{FP} give information on the relative sizes of $L,R$ needed for there to exist a matching of $L$ into $R$ w.h.p. 

We will prove the following theorem: it shows that the expected insertion time is $O(1)$, but only for a large value of $d$. The theorem focusses on the more interesting case where the load factor $n/m$ is close to one. When the load factor is small enough i.e. when $n\leq(1-\e)m$ the components of $\G$ will be bounded in expectation and so it is straightforward to show an $O(1)$ expected insertion time.
\begin{theorem}\label{th1}
Suppose that $n=(1-\e)m$ where $\e$ is a fixed positive constant, assumed to be small. Let $0<\th<1$ also be a fixed positive constant and let
\beq{defgamma}{
\g=5(1-\e)^{d/2}.
}
If $d^2\g\leq (1-\th)(d-1)$ then w.h.p. the structure of $\G$ is such that over the random choices in Steps 2,3,
\beq{eq1}{
\E(|P_k|)\leq 1+\frac{2}{\th}\text{ for }k=1,2,\ldots,n.}
Here $|P_k|$ is the length (number of edges) of $P_k$.
\end{theorem}
When $d$ is large the value of $\th$ in \eqref{eq1} is close to $d\e/2$. It can be seen from the proof that as $\e\to 0$, the value of $d$ needed is of the order $\bfrac{\log 1/\e}{\e}$. This is larger than the value $O(\log(1/\e))$ needed for there to be a perfect matching from $L$ to $R$ and finding an $O(1)$ bound on the expected insertion time for small $d$ remains as an open problem. We note that Theorem 1 of \cite{FPSS} allows small values of $d\geq 5+3\log1/\e$, but the BFS algorithm it relies on requires more space, shown to be $O(n^\d)$ extra space for constant $\d>0$, and is shown to have an expected insertion time of $d^{O(\log 1/\e)}$.

The problem here bears some relation to the {\em On-line bipartite matching problem} discussed for example in Chaudhuri, Daskalakis, Kleinberg and Lin \cite{CDKL}, Bosek, Leniowski, Sankowski and Zych \cite{BLSZ} and Gupta, Kumar and Stein \cite{GKS}. In these papers the bipartite graph is arbitrary and has a perfect matching and vertices on one side $A$ of the bipartition arrive in some order, along with their choice of neighbors in the other side $B$. As each new member of $A$ arrives, a current matching is updated via an augmenting path. The aim is to keep the sum of the lengths of the augmenting paths needed to be as small as possible. It is shown, among other things, in \cite{CDKL} that this sum can be bounded by $O(n\log n)$ in expectation and w.h.p. This requires finding a shortest augmenting path each time. Our result differs in that our graph is random and $|A|=(1-\e)|B|$ and we only require a matching of $A$ into $B$. On the other hand we obtain a sum of lengths of augmenting paths of order $O(n)$ in expectation via a random choice of path.
\section{Proof of Theorem \ref{th1}}
\subsection{Outline of the main ideas}
Let 
$$B_k=\set{v\in L_k: N(v)\cap \bar R_{k-1}=\emptyset}.$$
If $x\notin B_k$ in Step 2 of {\sc insert} then we will have found $P_k$.

Let $P=(x_1,\xi_1,x_2,\xi_2,\ldots,x_\ell)$ be a path in $\G$, where $x_1,x_2,\ldots,x_\ell\in L_{k-1}$ and $\xi_1,\xi_2,\ldots,\xi_{\ell-1}\in R_{k-1}$. We say that $P$ is {\em interesting} if $x_1,x_2,\ldots,x_\ell\in B_k$. We note that if the path $P_k=(x_1=v_n,\xi_1,x_2,\xi_2,\ldots,x_\ell,\xi_{\ell},x_{\ell+1},\xi_{\ell+1})$ then $Q_k=(x_1,\xi_1,x_2,\xi_2,\ldots,x_\ell)$ is interesting. Indeed, we must have $x_i\in B_k,\,1\leq i\leq\ell$, else {\sc insert} would have chosen $\xi_{i}\in \bar R_{k-1}$ and completed the round.  

Our strategy is simple. We show that w.h.p. there are relatively few long interesting paths and because our algorithm (usually) chooses a path at random, it is unlikely to be long and interesting. One caveat to this approach is that while all augmenting paths yield interesting sub-paths, the reverse is not the case. In which case, it would be better to estimate the number of possible long augmenting paths. The problem with this approach is that we then need to control the distribution of the matching $M_k$. This has been the difficulty up to now and we have avoided the problem by focussing on interesting paths. Of course, there is a cost in that $d$ is larger than one would like, but it is at least independent of $n$.

To bound the number of interesting paths, we bound $|B_k|$ and use this to bound the number of paths.
\subsection{Detailed proof}
Fix $1\leq k\leq n$. We observe that if $R_{k-1}=\set{y_1,y_2,\ldots,y_{k-1}}$ then 
\beq{yk}{
y_k \text{ is chosen uniformly from }\bar R_{k-1}
}
and is independent of the graph $\G_{k-1}$ induced by $L_{k-1}\cup R_{k-1}$. This is because we can expose $\G$ along with the algorithm. When we start the construction of $M_{k}$ we expose the neighbors of $v_{k}$ one by one. In this way we either determine that $S_k(v_{k})=\emptyset$ or we expose a uniformly random member of $S_k(v_{k})$ without revealing any more of $N(v_{k})$. In general, in Step 2, we have either exposed all the neighbors of $x$ and these will necessarily be in $R_{k-1}$. Or, we can proceed to expose the unexposed neighbors of $x$ until either (i) we determine that $S_k(x)=\emptyset$ and we choose a uniformly random member of $N(x)$ or (ii) we find a neighbor of $x$ that is a uniformly random member of $\bar R_{k-1}$. Thus 
\beq{Rkbarisrandom}{
R_{k-1}\text{ is a uniformly random subset of $R$.}
}

We need to show that $B_k$ is small. It is clear that $v_1\notin B_1$ i.e. $B_1=\emptyset$ and so we deal next with $2\leq k\leq d-2$. If $v_k\in B_k$ then $v_k$ must choose some vertex in $R$ three times. But,
\beq{2<k<d-1}{
\Pr(\exists 2\leq k\leq d-2,w\in R:v_k\text{ chooses $w$ three times})\leq (d-2)m\times m^{-3}=o(1). 
}
This implies that w.h.p. $B_k=\emptyset$ for $2\leq k\leq d-2$. We deal next with $d-1\leq k\leq n^{9/10}$. Since $N(v_k)$ is uniformly random, we see that
\beq{smallk}{
\Pr(\exists k\leq n^{9/10}:\,v_k\in B_k)\leq n^{1-d/10}=o(1)\text{ for }d>10.
}
Assume from now on that $n^{9/10}\leq k\leq n-1$. 
Let $\n_{k,\ell}$ denote the number of interesting paths with $2\ell - 1$ vertices. Let $\th,\g>0$ be as in the statement of Theorem \ref{th1}.
\begin{lemma}\label{lem1}
Given $A_0$ and $d$ sufficiently large,
\beq{large}{
\Pr\brac{\exists 2\leq \ell\leq A_0\log\log n:\n_{k,\ell}\geq (1+\th)k\g(d^2\g)^{\ell-1}}=o(n^{-2}).
}
\end{lemma}
The bound $o(n^{-2})$ is sufficient to deal with the insertion of $n$ items.

Before proving the lemma, we show how it can be used to prove Theorem \ref{th1}. We will need the following claims:
\begin{claim}\label{cl1a}
Let $\D$ denote the maximum degree in $\G$. Then for any $t\geq \log n$ we have $\Pr(\D\geq t)\leq e^{-t}$.
\end{claim}
{\bf Proof of Claim:}
If $v\in L$ then its degree $\deg(v)=d$. Now consider $w\in R$. Then for $t\geq \log n$,
\beq{deg1}{
\Pr(\exists w\in R: \deg(w)\geq t)\leq m\binom{dn}{t}\frac{1}{m^{t}}\leq m\bfrac{de}{t}^t\leq e^{-t}.
}\\
{\bf End of proof of Claim}
\begin{claim}\label{cl1}\ 
With probability $1-o(n^{-2})$, $\G$ contains at most $n^{1/2+o(1)}$ cycles of length at most $\Lambda=(\log\log n)^2$.
\end{claim}
{\bf Proof of Claim:}
Let $C$ denote the number of cycles of length at most $2\ell=\Lambda$. Then
$$\E(C)\leq\sum_{s=2}^\ell \binom{n}{s}\binom{m}{s}(s!)^2\bfrac{d}{m}^{2s}\leq \sum_{s=2}^\ell d^{2s}= n^{o(1)}.$$
Now let $C_1$ denote the number of cycles of length at most $\Lambda$ where we can only use the first $\min\set{\l,\deg(w)},\,\l=(\log n)^2$ edges incident with each vertex $w\in R$. Here ``first'' is defined in some canonical way. Then $C_1\leq C$ and Claim \ref{cl1a} implies that
\beq{new1}{
\Pr(C\neq C_1)\leq e^{-\l}=e^{-(\log n)^2}.
} 
Now $C_1$ depends on the $dn$ independent choices of edges in $\G$ and changing one choice of edge can only change $C_1$ by at most $\l^\Lambda$. Applying McDiarmid's inequality, and using $\E(C_1)\leq \E(C)$ we see that
$$\Pr(C_1\geq \E(C)+n^{1/2}\l^{2\Lambda})\leq \exp\set{-\frac{2n\l^{4\Lambda}}{dn\l^{2\Lambda}}}=o(n^{-2}).$$
Together with \eqref{new1}, this proves the claim.\\
{\bf End of proof of Claim}

These two claims imply the following:
\begin{multline}\label{cl1b}
\text{With probability $1-o(n^{-2})$ there are at most $n^{1/2+o(1)}(3\log n)^{2\ell}=n^{1/2+o(1)}$ vertices within}\\
\text{ distance at most $2A_0\log\log n$ of a cycle of length at most $\Lambda=(\log\log n)^2$.}
\end{multline}
Now let $p_{k,\ell}$ denote the probability that {\sc insert} requires at least $\ell$ steps to insert $v_{k}$. 

We finish the proof of the theorem by showing that 
\beq{proveth}{
\E(|P_k|)=1+2\sum_{\ell=2}^\infty p_{k,\ell}\leq 1+\frac{2}{\th}.
}
We observe that if $v_{k}$ has no neighbor in $\bar R_{k-1}$ and has no neighbor in a cycle of length at most $\Lambda$ then for some $\ell\leq A_0\log\log n$, the first $2\ell-1$ vertices of $P_n$ follow an interesting path. Hence, if $d^2\g\leq (1-\th)(d-1)$ then 
\begin{multline}\label{short}
\sum_{\ell=2}^{A_0\log\log n}p_{k,\ell}\leq O(n^{-1/2+o(1)})+ \sum_{\ell=2}^{A_0\log\log n}\frac{\n_{k,\ell}}{k(d-1)^\ell}\\
\leq O(n^{-1/2+o(1)})+(1 + \th)\sum_{\ell=2}^{A_0\log\log n}\frac{k\g(d^2\g)^{\ell-1}}{k(d-1)^\ell}\\
\leq o(1)+(1+\th)\sum_{\ell=2}^\infty(1-\th)^{\ell-1}=o(1)+\frac{1-\th^2}{\th}.
\end{multline}
{\bf Explanation of \eqref{short}:} Following \eqref{cl1b}, we find that the probability $v_k$ is within $2A_0\log\log n$ of a cycle of length at most $\Lambda$ is bounded by $n^{-1/2+o(1)}$. The $O(n^{-1/2+o(1)})$ term accounts for this and also absorbs the error probability in \eqref{large}. Failing this, we have divided the number of interesting paths of length $2\ell-1$ by the number of equally likely walks $k(d-1)^\ell$ that {\sc insert} could take. To obtain $k(d-1)^\ell$ we argue as follows. We carry out the following thought experiment. We run our walk for $\ell$ steps regardless. If we manage to choose $y\in \bar R_{k-1}$ then instead of stopping, we move to $v_k$ and continue. In this way there will in fact be $k(d-1)^\ell$ equally likely walks. In our thought experiment we choose one of these walks at random, whereas in the execution of the algorithm we only proceed as far the first time we reach $\bar R_{k-1}$. Finally, for the algorithm to take at least $\ell$ steps, it must choose an interesting path of length at least $2\ell-1$. 

Note next that 
\beq{LL}{
p_{k,A_0\log\log n}\leq O(n^{-1/2+o(1)})+3^{-A_0\log\log n}.
}
It follows that
\beq{LL1}{
\sum_{\ell=A_0\log\log n}^{(\log n)^{A_0}}p_{k,\ell}\leq \sum_{\ell=A_0\log\log n}^{(\log n)^{A_0}}p_{k,A_0\log\log n}=o(1).
}
We will use the result of \cite{FMM}: We phrase Claim 10 of that paper in our current terminology.
\begin{claim}\label{cl2}
There exists a constant $a>0$ such that for any $v\in L_{k-1}$, the expected time for {\sc insert} to reach $\bar R_{k-1}$ is $O((\log k)^a)$.
\end{claim}
It follows from Claim \ref{cl2} that for any integer $\r\geq 1$,
\beq{LL2}{
\Pr(|P_k|\geq \r(\log k)^{2a})\leq \frac{1}{(\log k)^{\r a}}.
}
Indeed, we just apply the Markov inequality every $(\log k)^{2a}$ steps to bound $|P_k|$ by a geometric random variable.

It follows from \eqref{LL2} that
\beq{LL3}{
\sum_{\ell\geq 3(\log k)^{2a}}p_{k,\ell}\leq \sum_{\r=3}^\infty\ \sum_{\ell/(\log k)^{2a}\in[\r,\r+1]}p_{k,\ell}\leq \sum_{\r=3}^\infty\frac{1}{(\log k)^{\r a-2a}}=o(1).
}
Theorem \ref{th1} now follows from \eqref{proveth}, \eqref{short}, \eqref{LL1} and \eqref{LL3}, if we take $A_0> 2a$.
\subsection{Proof of Lemma \ref{lem1}}
We will argue as in the proof of Claim \ref{cl1} that
\beq{nell}{
\Pr(\n_{k,\ell}\geq \E(\n_{k,\ell})+n^{3/4})\leq 2e^{-(\log n)^2}.
}
We let $\n_{k,\ell}^*$ be the number of interesting paths that only use the first $\min\set{\l,\deg(w)},\,\l=(\log n)^2$ edges incident with vertex $w\in R$. 
Then $\n_{k,\ell}^*\leq \n_{k,\ell}$ and Claim \ref{cl1a} implies that
\beq{new2}{
\Pr(\n_{k,\ell}\neq \n_{k,\ell}^*)\leq e^{-\l}=e^{-(\log n)^2}.
} 
Now $\n_{k,\ell}^*$ depends on the $dk$ independent choices of edges in $\G_k$ and changing one choice of edge can only change $\n_{k,\ell}^*$ by at most $\l^{2\ell}$. Applying McDiarmid's inequality, and using $\E(\n_{k,\ell}^*)\leq \E(\n_{k,\ell})$ we see that
$$\Pr(\n_{k,\ell}^*\geq \E(\n_{k,\ell})+n^{3/4})\leq \exp\set{-\frac{2n^{3/2}}{dk\l^{4\ell}}}\leq e^{-(\log n)^2}.$$
Together with \eqref{new2}, this proves \eqref{nell}.

It follows from \eqref{nell} that to finish the proof, all we need to show is that if $\th>0$ is an arbitrary positive constant
\beq{Expectation}{
\E(\n_{k,\ell})\leq (1+\th)k\g(d^2\g)^{\ell-1},
}
where $\g$ is as in \eqref{defgamma}.
\begin{claim}\label{Bsize}
Let
$$\cB_k=\set{|B_k|\geq k\g}.$$
Then
\beq{deg2}{
\Pr(\cB_k)=O(e^{-\Omega(n^{1/2})}).
}
\end{claim}
{\bf Proof of Claim:}\\
Let $B_{k,1}$ denote the set of vertices $v_i\in L_k$ such that round $i$ exposes at least $d/2$ edges incident with $v_i$. Then
$$\Pr(v_i\in B_{k,1})\leq (1-\e)^{d/2}.$$
It then follows from the Chernoff bounds that
\beq{show0}{
\Pr\brac{|B_{k,1}|\geq 2k(1-\e)^{d/2}}=O(e^{-\Omega(n^{1/2})}).
}
Next let 
$$B_{k,2}=\set{s\leq k:\text{ round $s$ does not end immediately in Step 2 with $x=v_s$.}}$$
Then, $\Pr(s\in B_{k,2})=\bfrac{s-1}{m}^d$ and this holds for each value of $s$ independently and so
$$\E(|B_{k,2}|)\leq \sum_{s=1}^k\bfrac{s-1}{m}^d\leq \frac{k^{d+1}}{(d+1)m^d}.$$
Now $|B_{k,2}|$ is the sum of independent $\set{0,1}$ random variables and so Hoeffding's theorem \cite{HOEF} implies that for a constant $\th>0$,
\beq{Ksize}{
\Pr\left(|B_{k,2}|\geq (1+\th)\frac{k^{d+1}}{(d+1)m^d}\right)=O(e^{-\e_1k})\text{ for some constant }\e_1=\e_1(d,\e,\th)>0.
}
Now if $B_{k,3}=\set{s\in B_k:\exists\ell\leq k, \ell\neq s\text{ s.t. round $\ell$ ends with $x=v_s$}}$ then $|B_{k,3}|\leq |B_{k,2}|$. Define $B_{k, 4} = B_k\setminus(B_{k, 1} \cup B_{k, 2} \cup B_{k, 3})$. Let $t > s$ be the first time that $v_s$ is re-visited by {\sc insert} or let $t=k$ if $v_s$ is not re-visited. Then $s\in B_{k, 4}$ only if in round $t$, at least $d/2$ unexposed edges incident to $s$ are found to be in $R_{t-1}$. It follows that
\beq{B4}{
E(|B_{k, 4}|) \leq k (1 - \e)^{d/2}.
}
Since membership of $s$ in $B_{k, 4}$ is determined by the random choices of $v_s$, $|B_{k, 4}|$ is the sum of independent random variables and so
\beq{show4}{
\Pr\brac{|B_{k,4}|\geq 2k(1-\e)^{d/2}}=O(e^{-\Omega(n^{1/2})}).
}
The claim follows from \eqref{show0}, \eqref{Ksize} and \eqref{show4}.\\
{\bf End of proof of Claim}

Given Claim \ref{Bsize}, we have 
\begin{align}
\E(\n_{k,\ell})&=\E(\n_{k,\ell}\mid \neg\cB_k)\Pr(\neg\cB_k) + \E(\n_{k,\ell}\mid \cB_k)\Pr(\cB_k)\\
&\leq k^{\ell} \g^\ell k^{\ell-1}\cdot \brac{(1+o(1))\frac{d}{k}}^{2\ell-2}+O(k^{2\ell -1}\cdot e^{-\Omega(n^{1/4})}),\label{lll}\\
&\leq(1+o(1))k\g(d^2\g)^{\ell-1}+o(1).
\end{align}
This proves \eqref{Expectation}.

{\bf Explanation of \eqref{lll}:} We can choose the vertex sequence  $\s=(x_1,\xi_1,\ldots,\xi_{\ell-1},x_\ell)$ of an interesting path $P$ in at most $|B_k|^{\ell} k^{\ell-1}$ ways, and we apply Claim \ref{Bsize}. Having chosen $\s$ we see that $((1+o(1))d/k)^{2\ell-2}$ bounds the probability that the edges of $P$ exist. To see this, condition on $\bar R_{k-1}$ and the random choices for vertices not on $P$. In particular, we can fix $R_{k-1}=\set{y_1,y_2,\ldots,y_{k-1}}$ from the beginning and this simply constrains the sequence of choices $y_1,y_2,\ldots,y_{k-1}$ to be a uniformly random permutation of $R_{k-1}$. Let $\cM_k$ be the property that $\G$ has a matching from $L_k$ to $R$. It is known that $\Pr(\cM_k)=1-O(n^{4-d})$. This will also be true conditional on the value of $\bar R_{k-1}$. This follows by symmetry. The conditional spaces will be isomorphic to each other. So for large $d$, we can assume that our conditioning is such that with probability $1-O(1/n^3)$ the edge choices by $x_1,x_2,\ldots,x_\ell$ are such that $\G_{k}$ has property $\cM_k$ with probability $1-O(n^{7-d})$. Recall from \eqref{yk} that the disposition of the edges of $\G_{k-1}$ is independent of $\bar R_{k-1}$. Now each edge adjacent to a given $x\in \s\cap L_k$ is a uniform choice over those edges consistent with $x$ being in $B_k$. But there will be at least $k-1$ such choices for such an $x$ viz. the vertices of $R_{k-1}$. Thus 
$$\Pr(P\text{ exists}\mid\cM_k)\leq \frac{\Pr(P\text{ exists})}{\Pr(\cM_k)}\leq (1+o(1))\bfrac{d}{k}^{2\ell-2}.$$
Note that $\Pr(\bar\cM_k)$ is only inflated by at most $\frac{1}{(1-\e)^{d\ell}}=o(n^{o(1)})$ if we condition on $x_1,x_2,\ldots,x_\ell$ making their choices in $\bar R_{k-1}$. This has to be compared with the unconditional probability of $O(n^{7-d})$.

This completes the proof of Theorem \ref{th1}.
\proofend
\begin{remark}
Along with an upper bound, we can prove a simple lower bound:
$$\E(|P_k|)\geq \frac{2}{1-(1-\e)^d}.$$
This follows from the fact that Step 2 of {\sc insert} ends the procedure with probability  $1-(1-\e)^{|S_k(x)|}$ and $|S_k(x)|\leq d$. 
\end{remark}
\section{Final Remarks}
There is plenty of room for improvement in the bounds on $d$ in Theorem \ref{th1}. It would be most interesting to prove an $O(1)$ bound on the expected insertion time for small $d$, e.g. $d=3,4,5$. This no doubt requires an understanding of the evolution of the matching $M$.

{\bf Acknowledgement:} We thank Wesley Pegden and the reviewers for their comments. We also thank Lutz Warnke for pointing out an error.

\end{document}